\documentclass[12pt,a4paper,superscriptaddress,preprint]{revtex4-1}
\usepackage[dvips]{graphicx}
\usepackage{amssymb,amsmath}

\begin{document}

\title{Totally asymmetric simple exclusion process \\ with a time-dependent boundary: \\ interaction between vehicles and pedestrians at intersections}
\date{\today}
\author{Hidetaka Ito}
\email{ito@jamology.rcast.u-tokyo.ac.jp}
\affiliation{%
Department of Aeronautics and Astronautics, School of Engineering, The University of Tokyo,\\
7-3-1 Hongo, Bunkyo-ku, Tokyo 113-8656, Japan
}%
\thanks{}%
\author{Katsuhiro Nishinari}%
\affiliation{%
Research Center for Advanced Science and Technology, The University of Tokyo,\\
4-6-1 Komaba, Meguro-ku, Tokyo 153-8904, Japan
}%

\begin{abstract}
Interaction between vehicles and pedestrians is seen in many areas such as crosswalks and intersections. In this paper, we study a totally asymmetric simple exclusion process with a bottleneck at a boundary caused by an interaction. Due to the time-dependent effect originating from the speed of pedestrians, the flow of the model varies even if the average hopping probability at the last site is the same. We analyze the phenomenon by using two types of approximations: (2+1)-cluster approximation and isolated rarefaction wave approximation. The approximate results capture intriguing features of the model. Moreover, we discuss the situation where vehicles turn right at the intersection by adding a traffic light at the boundary condition. The result suggests that pedestrian scrambles are valid to eliminate traffic congestion in the right turn lane. 
\end{abstract}

\maketitle

\section{INTRODUCTION}

Traffic congestion is one of major issues in the world \cite{chowdhury2000statistical,helbing2001traffic}. It is caused when the number of vehicles exceeds the limit of the traffic capacity. The phenomenon is inevitable provided we rely on vehicles and roads as a means of transportation. In recent years, traffic flow has been widely studied by physicists implementing a wide range of methods. Researchers have invented many models such as fluid \cite{kerner1993cluster,tomoeda2009new}, car following \cite{bando1995phenomenological,helbing1998generalized,jiang2001full,shamoto2011car} and cellular automaton (CA) \cite{kanai2005stochastic,nagel1992cellular,sakai2006new}. The totally asymmetric simple exclusion process (TASEP) is one of the most successful CA models for the analysis of phenomena far from equilibrium such as traffic flow \cite{blythe2007nonequilibrium,chou2011non}. We consider that the road is divided into $n$ discrete lattice sites, where each site is occupied or empty. Each vehicle hops with probability $p$ if the front site is empty. Vehicles enter with probability $\alpha$ at one boundary and exit the system with probability $\beta$ at the other boundary. It is known that the flow of the TASEP depends on the updating schemes \cite{rajewsky1998asymmetric}. In random updating, the lattice site to be updated is decided randomly. In fully parallel updating, all lattice sites are updated simultaneously. The flow is larger for fully parallel updating than that for random updating. The TASEP with fully parallel updating is identical to the Nagel--chreckenberg (NS) model with $V_{max}=1$ \cite{nagel1992cellular}. It is well known that the NS model is one of the realistic CA models of traffic flow. Even though the rules of CA models are quite simple, because of exclusion and many-body interactions, they have intriguing features such as phase transition from free flow to traffic jam, which have attracted many physicists.

Traffic lights are a major cause of traffic jams since they completely block the flow of vehicles. The other method for controlling traffic flow at an intersection has been proposed: roundabouts \cite{fouladvand2004characteristics,PhysRevE.73.036101}. These allow vehicles to pass an intersection without waiting when there are few vehicles near them. Due to this advantage, roundabouts are used in many countries. However, they have difficulty in managing heavy traffic, hence, traffic lights remain the main devices used to control vehicles under such conditions. Since the traffic light was invented in 1868, better ways of it has been investigated enthusiastically \cite{brockfeld2001optimizing,PhysRevE.70.016107,varas2009resonance}. Popkov {\it et al.} investigated the TASEP with a traffic light at the boundary \cite{popkov2008asymmetric}. Using mean field theory, he derived inviscid Burgers' equation and studied it. The simulation results and numerical integration of the equation show a stationary sawtooth structure of the density. Although the exact stationary state of the TASEP with ordinary open boundaries has already been derived \cite{evans1999exact,de1999exact}, the modification of the boundary of the TASEP such as a traffic light boundary, makes it difficult to analyze the features of traffic flow. Recently, the TASEP with varying boundary conditions has been studied. Woelki studied the TASEP whose entry probability, $\alpha$, depends on the number of particles in the system \cite{PhysRevE.87.062818}. Analytic results using mean field theory are in agreement with the simulation results. In the context of biology, Wood investigated the TASEP with gate boundaries \cite{wood2009totally}. Each particle exits the system with a receptor, which behaves like a gate. These two models adopted random updating; the TASEP with fully parallel updating is little studied so far. 

There is another main reason for traffic congestion: interaction between vehicles and pedestrians at an intersection. Vehicles turn right (left in some countries) and pedestrians pass the crossing at the same time. This reduces the number of vehicles that can pass the intersection, and sometimes it causes traffic accidents. The interaction between vehicles and pedestrians is of interest among researchers because of its large impact on society; it is analyzed using continuous models \cite{helbing2005analytical,jin2013dynamic} and CA models \cite{li2009modeling}. Moreover, a new type of signal has been proposed to tackle the issue as a practical method: the {\em pedestrian scramble (diagonal crossing)}. The pedestrian scramble is one type of a traffic signal phasing scheme that allows vehicles and pedestrians to separately pass a crossing by segregating the time pedestrians and vehicles pass. When vehicles pass the crossing, all pedestrians have to wait. Then, pedestrians traveling in all directions can pass the crossing. Traffic signals that are operated using this scheme are called {\em pedestrian--vehicle separation signals}. The pedestrian scramble at Hachiko Square in Shibuya, Tokyo, is a famous example of a pedestrian scramble. The pedestrian scramble was invented in the US, and recently, it has become popular because it is safer for pedestrians. A large number of pedestrian--vehicle separation signals have been implemented not only in Japan but other countries. However, the traffic capacity of neither pedestrian--vehicle separation signals nor ordinary signals is well understood. Therefore, in this paper, we discuss the traffic capacity of the right (left) turn lane at an intersection. We develop the TASEP with open boundaries, and the exit is connected to the crossing. We propose three models: intersection without traffic lights, with an ordinary traffic light, and with a pedestrian--vehicle separation signal. First, we consider the situation of the intersection without traffic lights, for simplicity. It is equivalent to vehicles that attempt to pass crosswalks. Then, we consider a traffic light at the intersection. Furthermore, we compare the traffic capacity of pedestrian--vehicle separation signals and ordinary signals. 

This paper is organized as follows: In Sec.~\ref{sec:2}, we introduce our model without traffic lights and explain the rules for both pedestrians and vehicles. In Sec.~\ref{sec:3}, we analyze the pedestrian flow and its effect on traffic flow. In Sec.~\ref{sec:4}, we discuss the results of Monte Carlo simulations. In Sec.~\ref{sec:5}, we discuss the limiting cases and derives two types of approximations: (2+1)-cluster approximation (TCA) and isolated rarefaction wave approximation (IRA). In Sec.~\ref{sec:6}, we study the model with a traffic light by using both simulations and theories. Additionally, we compare the traffic capacity for pedestrian--vehicle separation signals and ordinary signals. Finally, Sec.~\ref{sec:7} is devoted to our concluding discussions.

\begin{figure}
\begin{center}
\includegraphics[width=18cm]{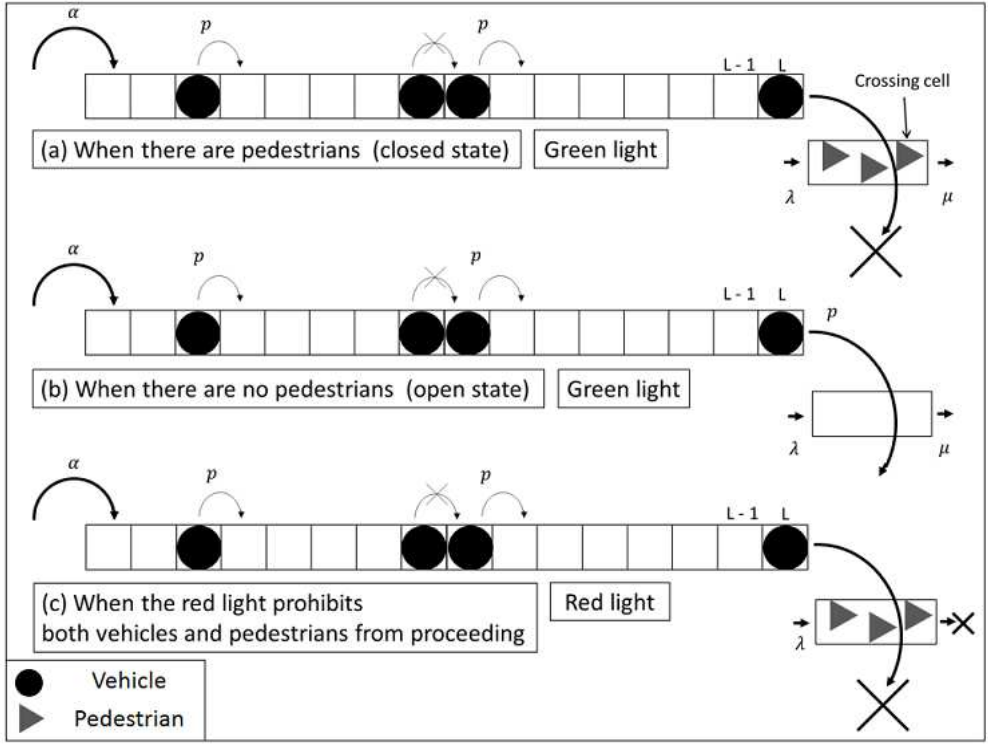}
\end{center}
\caption{Schematic view and update procedure for the model. The crossing cell for pedestrians is located at the right corner and the triangular particles are pedestrians. The number of pedestrians who go into the system at each time step is distributed under the Poisson distribution with parameter $\lambda$. Each pedestrian exits the system with probability $\mu$. Other cells are the road for vehicles, which are represented by the round particles. Vehicles enter the left side of the road with probability $\alpha$ and hop with probability $p$. (a) When there are pedestrians, no vehicles can exit the system. (b) When there are no pedestrians, vehicles are allowed to pass the crossing with probability $p$. (c) In Sec.~\ref{sec:6}, a traffic light is added to the model. When the traffic light is red, neither vehicles nor pedestrians cannot exit.}
\label{fig:13}
\end{figure}

\section{TASEP CONNECTED TO A PEDESTRIAN CROSSING}
\label{sec:2}

We consider the TASEP, which represents interactions between pedestrians on the crossing and vehicles on the road. The right boundary represents the intersection where cars turn right and pedestrians enter the crossing. Figure~\ref{fig:13}(a) and \ref{fig:13}(b) show the schematic view of the model during the green light period. They also represent the case without traffic lights. Figure~\ref{fig:13}(c) is the case that appears only if there is a traffic light: the red light period. The direction of vehicles and pedestrians is the same, and vehicles near the right boundary turn right at the intersection. We separately consider two models: for pedestrians and vehicles. We combine these models to represent interactions. We adopt the updating rule that is fully parallel. In this section, no traffic lights are implemented. 

\subsection{Pedestrians}

There is one cell for pedestrians at the right end of the model as shown in Fig.~\ref{fig:13}. It symbolizes the crossing and is defined as the {\em crossing cell}. The crossing cell is sufficiently large compared to the size of pedestrians. Thus we assume that it can contain an infinite number of pedestrians. We also assume that the arrival of pedestrians is independent of other pedestrians. Therefore, it is suitable to use the Poisson distribution. The number of pedestrians who enter the crossing cell at each time step is distributed under the Poisson distribution with parameter $\lambda$, where $\lambda$ is the arrival rate for pedestrians. They exit the crossing cell taking several time steps, and the time each pedestrian takes varies. Thus, we define each pedestrian as exiting the crossing cell with probability $\mu$, which denotes the speed of pedestrians passing the crossing cell. It is also possible to consider bicycles entering the crossing cell instead of pedestrians. In the case of bicycles, we take a larger $\mu$. In the case of elderly pedestrians who cannot to walk fast, $\mu$ is small. The model is similar to an M/M/$\infty$ queue; however, our model is a discrete-time system.

\subsection{Vehicles}
 
The model for vehicles is the TASEP with open boundaries, defined on a one-dimensional lattice with $L$ cells that are labeled $i=1,\dots,L$. Each cell is empty or occupied by a particle. Two particles cannot share one cell because of the exclusive property. At each time step, vehicles jump to the next cell with probability $p$ if there are no vehicles occupying that cell. If the boundary cell, $i=1$, is not occupied, the vehicle can enter the cell with probability $\alpha$. The updating rules for vehicles on cells $i=1,\dots,L-1$ are unaffected by pedestrians. However, vehicles on the boundary cell, $i=L$, interact with pedestrians. This site represents the road in front of the crossing cell. Here the correlation between the boundary cell and pedestrians is described. According to the law in Japan, for safety, drivers must wait if they notice pedestrians who want to pass the crossing. Vehicles are not permitted to enter the crossing cell, even if it contains only one pedestrian. Additionally, since the area that pedestrians can walk in is small enough for vehicles to pass the crossing cell immediately after they enter it, it is appropriate to assume that vehicles do not enter the crossing cell. If there are no pedestrians in the crossing cell, the vehicle on the cell $i=L$ jumps outside the system with probability $p$, which is the same as the hopping probability in the bulk cells, $i=1,\dots,L-1$. The parameter, $\beta(t)$, is dependent on time and pedestrians' behavior. It seems to be quite simple, but it leads to intriguing many-body interactions as shown later.

\section{Analysis of Pedestrians' behavior and the effect on vehicles} 
\label{sec:3}

Since we assume that the behavior of pedestrians is not affected by vehicles, it can be analyzed independently of vehicles' behavior. Let $\pi_{j}$ denote the probability of $n$ pedestrians in the site. The probability $\pi_{n}$ is determined by 
\begin{equation} 
\pi_{n} = \frac{1}{n!} \left( \frac{\lambda}{\mu}\right)^n e^{-\frac{\lambda}{\mu}}. \hspace{15pt} (n \ge 0)
\end{equation}
The derivation of the equation is given in the Appendix. For vehicles' behavior, it only matters whether there is at least one pedestrian in the site. We call the state of the crossing cell the {\em open state} if the crossing cell is empty, and the {\em closed state} if there is at least one pedestrian occupying the cell. Let $P^{O}$ and $P^{C}$ denote the probabilities of the open state and closed state, respectively. These are given by $P^{O}= e^{-\lambda/\mu}$ and $P^{C}=1-e^{-\lambda/\mu}$. Although $\beta(t)$ depends on time, we obtain the mean hopping probability 
\begin{equation} 
\bar{\beta} = p e^{-\lambda/\mu}.
\end{equation}
The parameter $\bar{\beta}$ corresponds to the probability that the particle at the cite $i=L$ exits of the TASEP with ordinary open boundaries, $\beta$. The parameter $\lambda$ should be variable and $\mu$ should be fixed in this context. Since $\lambda$ and $\bar{\beta}$ have the relation, $\lambda$ can be expressed as a function of $\bar{\beta}$. For the analysis of the ordinary TASEP model, $\alpha$ and $\beta$ ($=\bar{\beta}$) are the main parameters. Therefore, $\bar{\beta}$ is used as an indicator of the number of pedestrians instead of $\lambda$. The transition matrix for the pedestrian model can be obtained as follows: We define 
\[
T_{ped}=\left(
    \begin{array}{cc}
      T^{O;O} &T^{O;C}  \\
      T^{C;O} &T^{C;C} \\
    \end{array}
  \right)
\]
as the transition probability of the state of the crossing cell, where $T^{A;B}$ denotes the probability of A after updating if the state before updating is B. The states $O$ and $C$ are open state and closed state, respectively. The probability that there are pedestrians before updating and the next state is the open state, $T^{O;C}$, is $\sum_{n=1}^{\infty} \pi_n P^{0;n}$, where $P^{0;n}$ is the probability of $n$ pedestrians after updating when there are $n$ pedestrians before updating. Because the probability of the closed state is $\sum_{n=1}^{\infty} \pi_n$, we obtain the conditional probability
\begin{equation} 
T^{O;C} = \frac{\sum_{n=1}^{\infty} \pi_{n} P^{0,n}}{\sum_{n=1}^{\infty} \pi_{n}} = \frac{e^{-\frac{\lambda}{\mu}}(1-e^{-\lambda})}{1-e^{-\frac{\lambda}{\mu}}}.
\end{equation} 
In this way, the transition matrix $T_{ped}$ is given by
\begin{eqnarray}
T_{ped}=\left(
    \begin{array}{cc}
e^{-\lambda} & \frac{e^{-\frac{\lambda}{\mu}}(1-e^{-\lambda})}{1-e^{-\frac{\lambda}{\mu}}} \\
1 - e^{-\lambda} & 1 - \frac{e^{-\frac{\lambda}{\mu}}(1-e^{-\lambda})}{1-e^{-\frac{\lambda}{\mu}}} \\
    \end{array}
  \right).
\end{eqnarray}
Although the mean probability of the open state depends only on $\lambda/\mu$, the transition probabilities do not. If we keep the mean probability of the open state constant, then $\mu$ denotes the degree of independence of the next state of the crossing cell with respect to the present state. Large $\mu$ indicates that the next state is almost independent of the present state. That is, the fact that pedestrians entering the crossing cell can quickly pass indicates that new pedestrians entering the crossing cell can significantly influence the next state of the cell. In the case of a small $\mu$, there is a strong tendency for the crossing cell to keep its present state. Once the crossing cell opens, it remains open for a long time and vice versa. This is because each pedestrian takes a much longer time to finish passing the crossing cell. Additionally, $\lambda$ is also small, which indicates a low arrival rate of pedestrians compared to the case of large $\mu$. This behavior at the boundary is similar to the traffic light boundary in terms of the length of the open state (green) time and closed state (red) time; although green duration and red duration at the boundary are not constant.

\section{SIMULATION RESULTS}
\label{sec:4}

\begin{figure}
\begin{center}
\includegraphics[width=9cm]{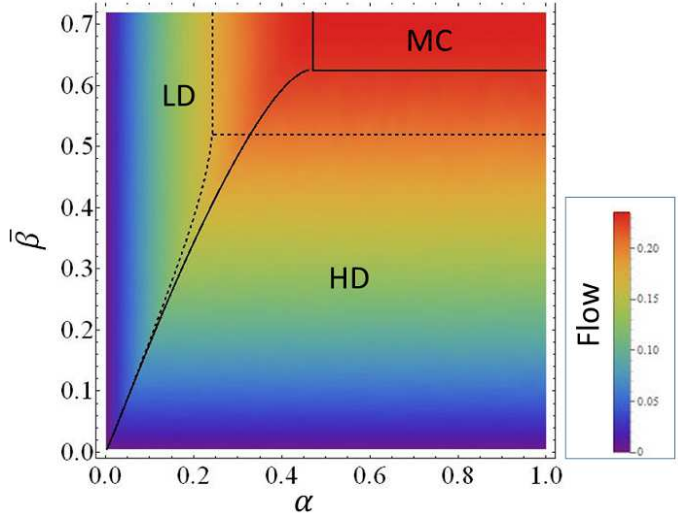}
\end{center}
\caption{(Color online) Phase diagram. Colors represents the value of the average flow for the parameters $p=0.72$ and $\mu=0.1$. The vertical axis is not $\beta$ but the average exit probability $\bar{\beta}$. The conditions $\bar{\beta}=p$ and $\bar{\beta}=0$ correspond to $\lambda=0$ and $\lambda \rightarrow \infty$, respectively, since $\bar{\beta} = p e^{-\lambda/\mu}$. The dashed and solid lines represent border lines obtained by (2+1)-cluster approximation (TCA) and isolated rarefaction wave approximation (IRA). For these parameters, the result of the IRA capture the essence of the model.}
\label{fig:4}
\end{figure}

\begin{figure*}
\begin{center}
\includegraphics[width=18cm]{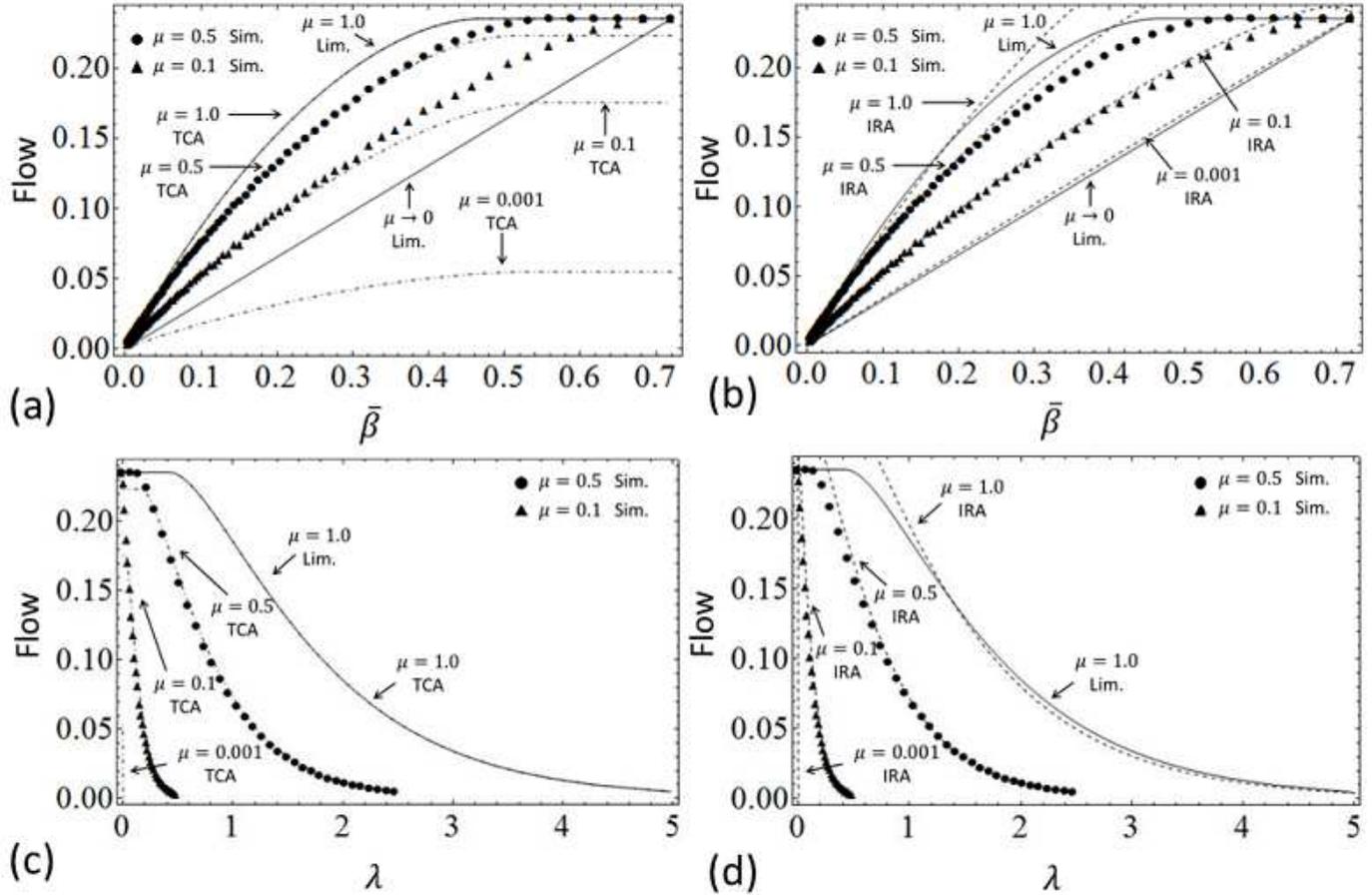}
\end{center}
\caption{(a) Average flow plotted against the average exit probability $\bar{\beta}$ for the parameters $p=0.72$, and $\alpha=1$. The figure represents the traffic capacity for the case. The circles and triangles correspond to $\mu=0.5$ and $\mu=0.1$, respectively. The solid lines show the result of the theoretical analysis of two limiting cases, $\mu=1$ and $\mu \rightarrow 0$. The dot-dashed lines represent the border-lines obtained by the TCA. For $\mu = 1$, the result of the TCA and the limiting case coincide exactly. The result of the TCA is in agreement with the simulation when $\mu$ is large. (b) The dots and solid lines are the same as (a). The dashed lines represent the border-lines obtained by the IRA. The result of the IRA is in agreement with the simulation when $\mu$ is small. Although (c) and (d) show the same data as (a) and (b), respectively, the horizontal axes of (c) and (d) are $\lambda$.}
\label{fig:2}
\end{figure*}

\begin{figure}
\begin{center}
\includegraphics[width=8cm]{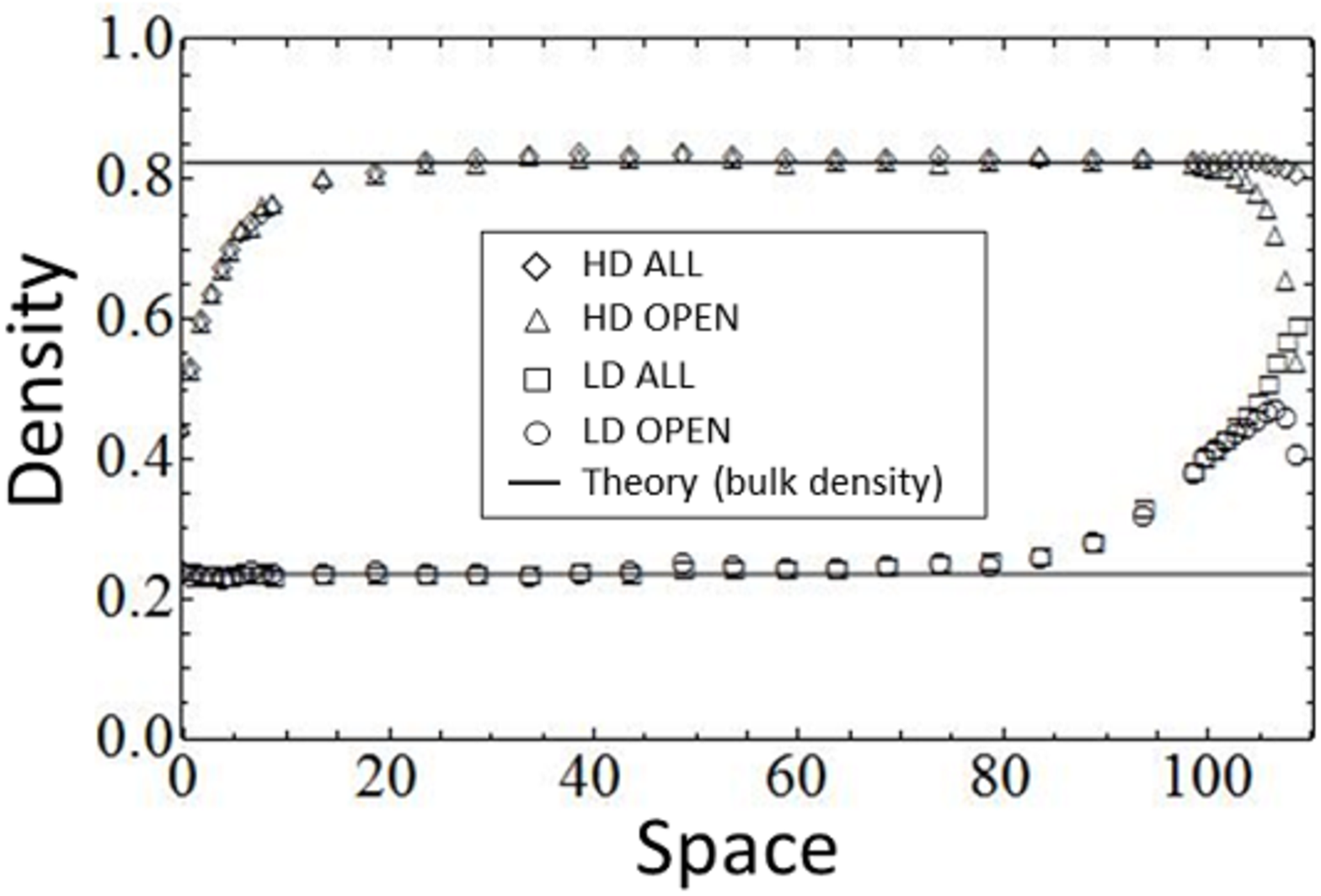}
\end{center}
\caption{Density plotted against the cell site. In this figure, we adopt $L=110$ to better illustrate the boundary effect. The diamonds and triangles correspond to the data in the HD phase ($\alpha=0.2$, $\lambda=0.24$, $\mu=0.2$, $\bar{\beta}= 0.21\ldots$), and the squares and circles correspond to the data in the LD phase ($\alpha=0.2$, $\lambda=0.12$, $\mu=0.2$, $\bar{\beta}= 0.39\ldots$). The diamonds and squares are average densities of all time steps. In contrast, the triangles and circles are average densities of time steps at which the state of the pedestrians is the open state. The solid lines are the approximation results of the bulk density obtained using the IRA.}
\label{fig:10}
\end{figure}

Now, we discuss various simulation results. In the following, we set $p=0.72$ and the number of sites is $2 \times 10^3$. We perform simulations for $5.0 \times 10^5$ time steps for each situation. The simulation results for flow and density are averaged over the last $2.5 \times 10^5$ steps. The phase diagram for vehicles is shown in Fig.~\ref{fig:4}. The dashed and solid lines are borders of these phases obtained using approximations discussed in Sec.~\ref{sec:5}. It is known that the TASEP has three phases; low-density (LD), high-density (HD) and maximum current (MC). The LD and HD phases correspond to free flow and jamming flow, respectively. Likewise, the model has three phases for all $\mu$. However, the borders of these phases are different in their locations. The co-existence line is the border between the LD and HD phases. The co-existence line is important because it represents the border line between free flow and jamming states. As $\mu$ decreases, the co-existence line moves to the upper side, which means the jamming area becomes wide. Moreover, we investigate the flow in the case $\alpha=1$, which explicitly shows the traffic capacity of traffic flow. In this case, only the HD and MC phases are realized. The traffic capacity plotted against $\bar{\beta}$ is shown in Fig.~\ref{fig:2}(a) and \ref{fig:2}(b). Even though the total time steps of the open state are the same, the traffic capacity depends on $\mu$. Additionally, the traffic capacity plotted against $\lambda$ instead of $\bar{\beta}$ is shown in Fig.~\ref{fig:2}(c) and \ref{fig:2}(d). We see that the flow exponentially drops with respect to $\lambda$, which shows that even a few pedestrians can reduce the traffic capacity. Figure~\ref{fig:10} shows the density diagram plotted against the site location. The density diagram is similar to the ordinary TASEP. However, there is a subtle difference between TASEP with ordinary boundaries and the model we study. If we decompose $\beta$ of ordinary TASEP into the probability that vehicles at the site $L$ hop, which is $p$, and the probability that the crossing cell is the open state $P^{O}$, we can treat the ordinary TASEP like the model. In this sense, the density of the ordinary TASEP is independent of the present state of the crossing cell since $\beta$ is constant. In contrast, the model is dependent on the present state of the crossing cell. The model has a strong tendency to keep the present crossing cell's state. Once the crossing cell opens, it remains open; this in turn decreases the density calculated near the crossing cell provided the previous state of the crossing cell is open. Additionally, the density of the model near the crossing cell is lower than the bulk density in the HD phase. These effects are large when $\mu$ is small. These phenomena are peculiar to the model and show that many vehicles are correlated near the crossing cell because of the property of that cell.

\section{THEORETICAL ANALYSIS}
\label{sec:5}

Let us consider the model taking the thermodynamic limit ($L \to \infty$). If the phase is LD or MC, the resulting bulk density and flow are independent of the site $L$ as seen in the previous section. In this case, the flow of the LD and MC phases are known to be  $J_{LD}=\alpha (p-\alpha)/(p-\alpha^2)$ and $J_{MC}=(1-\sqrt{1-p})/2$, respectively. In the case of the HD phase, the time-dependent leaving probability $\beta(t)$ plays an important role, and the flow does not depend only on the average exit probability $\bar{\beta}$ in the jamming case. The exclusion rule of the TASEP leads to complex n-body interactions which is unique to the model with time-dependent-$\beta(t)$.

\subsection{Limiting cases}

\begin{figure*}
\begin{center}
\includegraphics[width=14cm]{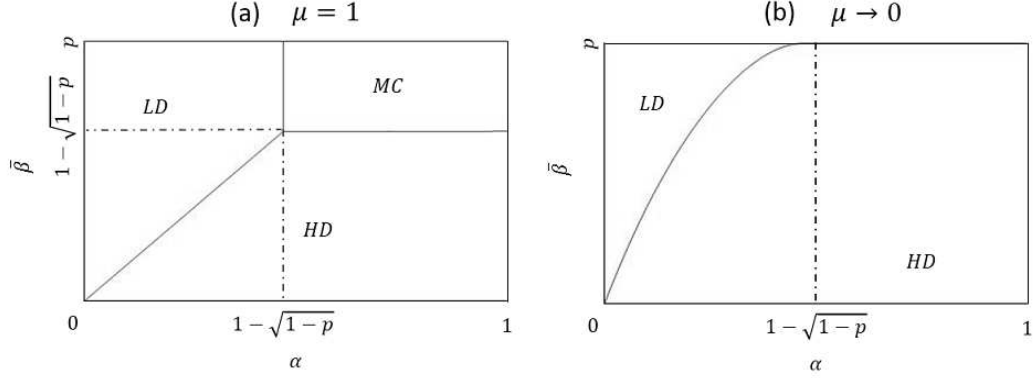}
\end{center}
\caption{Phase diagrams of two limiting cases. (a) When $\mu=1$, the diagram is the same as the TASEP with ordinary open boundaries. As $\mu$ decreases, the triple point approaches large $\bar{\beta}$. (b) When $\mu \rightarrow 0$, the MC phase vanishes.}
\label{fig:5}
\end{figure*}

If $\mu=1$, the model become the TASEP with ordinary boundaries for parameters $\alpha$ and $\bar{\beta}$. The next state of the boundary is independent of the previous state i.e., $\bar{\beta} = \beta$. This is because all pedestrians in the crossing cell always pass the crossing cell, and the state of the crossing cell depends only on whether new pedestrians enter. Thus, the flow of the HD phase is 
\begin{equation}
J_{HD} = \frac{\bar{\beta} (p-\bar{\beta})}{p-\bar{\beta}^2} \hspace{15pt} (\mu = 1).
\end{equation}
The co-existence line is a straight line, $\bar{\beta} = \alpha, \alpha \le 1- \sqrt{1-p}$, as shown in Fig.~\ref{fig:5}(a). 

In the case of $\mu \rightarrow 0$, with $\bar{\beta}$ constant, $T_{ped}$ becomes the $2 \times 2$ identity matrix. Even one person blocks the crossing cell for infinite time steps, while pedestrians arriving rate approaches $0$. As $\bar{\beta}$ is finite, the ratio of time for the open and closed states is finite, while one state is endless. For the case of the HD phase, most sites near the crossing cell are occupied by vehicles when the crossing cell opens after a prolonged closed state. At first, the cluster of vehicles passes the crossing cell, which leads to a large flow. Then, the number of vehicles near the crossing cell approaches the steady state after prolonged interactions and the density of the steady state approaches the density of the MC phase $1/2$. Very small $\mu$ allows the crossing cell to open for such a long time that the density near the crossing cell and the flow converge to $1/2$ and $(1-\sqrt{1-p})/2$, respectively. Although it takes time for the transition from the initial HD phase to the MC phase, the difference of flow while transferring can be neglected since the transition time is infinitesimal considering that the crossing cell keeps an open state for infinite time steps. Thus, vehicles' outflow is simply given by $(1-\sqrt{1-p})/2$ per unit time step during the open state. The crossing cell opens for $e^{-\lambda/\mu}=\bar{\beta}/p$ step per unit time step. As a result, the average flow converges 
\begin{equation}
\label{eq:mu0}
J_{HD} \rightarrow \frac{1-\sqrt{1-p}}{2} \frac{\bar{\beta}}{p} \hspace{15pt} (\mu \rightarrow 0).
\end{equation} 
In this case, the area of the MC phase in the phase diagram vanishes and the MC phase is realized only when $\bar{\beta} = p$. The co-existence line is a curve line given by
\begin{equation}
\bar{\beta} = \frac{2p}{1-\sqrt{1-p}}\frac{\alpha (p-\alpha)}{p-\alpha^2}, \hspace{15pt} \alpha \le 1- \sqrt{1-p}.
\end{equation}
The flow is always lower than the TASEP with ordinary boundaries for all $\bar{\beta}$. The phase diagram for cases of $\mu \to 0$ is shown in Fig.~\ref{fig:5}(b). 

In other cases, the flow is in between the cases for $\mu=1$ and $\mu \to 0$. Large $\mu$ allows the cycle of open and closed states to repeat quickly. The flow immediately after the crossing cell opens is large because of the cluster of vehicles that are accumulated near the crossing cell when it is in the closed state, which leads to the large flow. In contrast, small $\mu$ allows vehicles at the site, $L$, to hop for a long time. The flow is small after the cluster of packed vehicles passes the crossing. Thus, as $\mu$ decreases, the flow decreases. For all these reasons, even if the amount of time when the crossing cell is occupied by pedestrians is the same, the flow is lower in the case of small $\mu$. Pedestrians affect the flow of vehicles much more than bicycles.

\subsection{The (2+1)-cluster approximation} 

\begin{figure}
\begin{center}
\includegraphics[width=8.8cm]{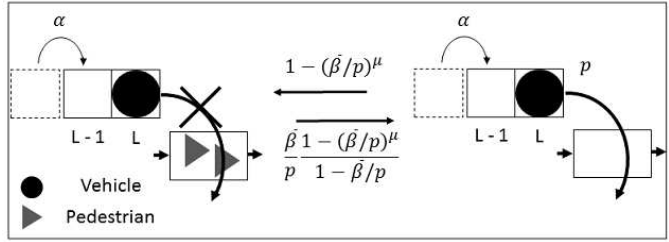}
\end{center}
\caption{Schematic representation of the TCA. Three cells at the right boundary and the cell for pedestrians are illustrated. Upper and lower figures represent the cases of closed state and open state, respectively. Additionally, the state of the crossing cell changes with the probability shown in the figure. The result captures features of the model when $\mu$ is large. When $\mu=1$, the approximation result is exact.}
\label{fig:14}
\end{figure}

It is known that the flow of the TASEP with ordinary open boundaries in the LD and HD phases can be exactly obtained using the two cluster approximation. However, an ordinal two-cluster approximation is not appropriate since it neglects correlation between vehicles and pedestrians. We therefore develop the microscopic {\em (2+1)-cluster approximation} (TCA). A schematic representation of the approximation method is shown in Fig.~\ref{fig:14}. The numbers ``$2$'' and ``$1$'' represent the last two cells and the crossing cell, respectively. The sites $L-1$, $L$, and the crossing cell are investigated in this approximation method. The stationary state of 2+1 cells is denoted by
\begin{equation} 
\Phi= {}^t\!(P_{0,0}^{O},P_{0,1}^{O},P_{1,0}^{O},P_{1,1}^{O},P_{0,0}^{C},P_{0,1}^{C},P_{1,0}^{C},P_{1,1}^{C}),
\end{equation}
where the first subscript, second subscript, and superscript are the states of the site at $L-1$, $L$, and the crossing cell, respectively. The transition matrix is defined by $T$, with size $8 \times 8$. $\Phi$ is constructed as follows: First, we construct the transition matrix of the model $T$ using the probability of injection to the site $L-1$, which is denoted by $\alpha_2$. The transition matrix of the model T is given by
\begin{eqnarray}
T= 
\left(
    \begin{array}{cc}
     (\frac{\bar{\beta}}{p})^\mu T_2(\alpha_2,p) & \frac{\bar{\beta}}{p} \frac{1- (\frac{\bar{\beta}}{p})^\mu}{1- \frac{\bar{\beta}}{p}} T_2(\alpha_2,0)   \\
       \left(1-(\frac{\bar{\beta}}{p})^\mu\right) T_2(\alpha_2,p) & \left(1- \frac{\bar{\beta}}{p} \frac{1- (\frac{\bar{\beta}}{p})^\mu}{1- \frac{\bar{\beta}}{p}} \right) T_2(\alpha_2,0) \nonumber
    \end{array}
  \right),
\end{eqnarray}
where $T_2(\alpha_2,p)$ and $T_2(\alpha_2,0)$ are the transition matrices of the TASEP with two lattice sites and ordinary open boundaries $T_2(\alpha,\beta)$ in cases of $\alpha=\alpha_2 \land \beta = p$ and $\alpha=\alpha_2 \land \beta = 0$. The $4 \times 4$ matrix $T_2(\alpha,\beta)$ is given by
\begin{eqnarray}
T_2(\beta)&=&
\left(
\begin{array}{cccc}
 1-\alpha  & (1-\alpha ) \beta  & 0 & 0 \\
 0 & (1-\alpha ) (1-\beta ) & \beta & 0 \\
 \alpha  & \alpha  \beta  & 1-\beta & \beta  \\
 0 & \alpha  (1-\beta ) & 0 & 1-\beta  \\
\end{array}
\right).
\end{eqnarray}
Then, we construct the stationary state $\Phi$, which is given as the solution of the equation
\begin{equation} 
\Phi=T \Phi
\end{equation}
with the normalization condition. We calculate $\alpha_2$, since $\alpha_2$ is currently unknown. An approximation is necessary to calculate $\alpha_2$ because the density at the site $L-2$ is not yet obtained. Since the difference in density of adjacent cells is not large, the density of the sites $L-1$ and $L$ is used instead of that of the sites $\L-2$ and $\L-1$ to obtain $\alpha_2$. The probability $\alpha_2$ is given as one of the solutions of the following equation:
\begin{equation} 
\alpha_2 = p\frac{P_{1,0}^{O}+P_{1,0}^{C}}{P_{0,0}^{O}+P_{1,0}^{O}+P_{0,0}^{C}+P_{1,0}^{C}}.
\end{equation}
We select the largest solution that satisfies $0 \le \alpha_2 \le 1$. Finally, substituting the solution of $\alpha_2$, the flow is given by
\begin{equation} 
J_{HD} = p (P_{0,1}^{O}+P_{1,1}^{O}).
\end{equation}
Because $J_{HD}$ has only one maximum value, let $\bar{\beta}_{max}$ and $J_{HDmax}$ denote $\bar{\beta}$ and $J_{HD}$ when $J_{HD}$ is a maximum. In the case of the TASEP with ordinary boundaries, $\bar{\beta}_{max}$ is the border between the HD and MC phases. The model is in the HD phase only when $\bar{\beta} < \bar{\beta}_{max}$ and $J$ keeps $J_{HDmax}$ when $\bar{\beta} \ge \bar{\beta}_{max}$ (MC phase). Even in this case, we assume that the TCA result of $J_{HD}$ is valid in the range $\bar{\beta} < \bar{\beta}_{max}$. 

We compute other variables. Since the flow in the HD and LD phases is the same on the co-existence line, this line is given as a solution of the equation $J_{HD}=J_{LD}$. Furthermore, the bulk density is given as the larger solution of the cubic equation
\begin{equation} 
J_{HD}=\frac{1-\sqrt{1-4p\rho_{bulk}(1-\rho_{bulk})}}{2}
\end{equation}
by using the conservation law for the flow. The approximation results are shown in Figs.~\ref{fig:4}, \ref{fig:2}, and \ref{fig:7}(a). The theoretical predictions agree well with the simulation results when $\mu$ is large and $\bar{\beta}$ is small. If $\mu$ is small or $\bar{\beta}$ is large, the flow obtained by the approximation is smaller than the simulation results. In this case, neglecting interactions of over two vehicles is not sufficient, although the TCA result of the TASEP with ordinary boundaries is exact. Therefore, we need other approximation methods to capture features when $\mu$ is small.

\subsection{The isolated rarefaction wave approximation}

\begin{figure}
\begin{center}
\includegraphics[width=8cm]{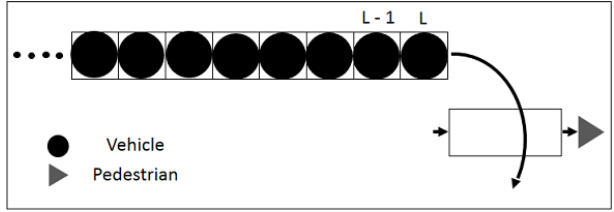}
\end{center}
\caption{Schematic representation of the IRA. Whenever the last pedestrian exit the system, we assume that the road near the crossing cell is full of vehicles. The result capture features of the model when $\mu$ is small.}
\label{fig:15}
\end{figure}

\begin{figure}
\begin{center}
\includegraphics[width=8.5cm]{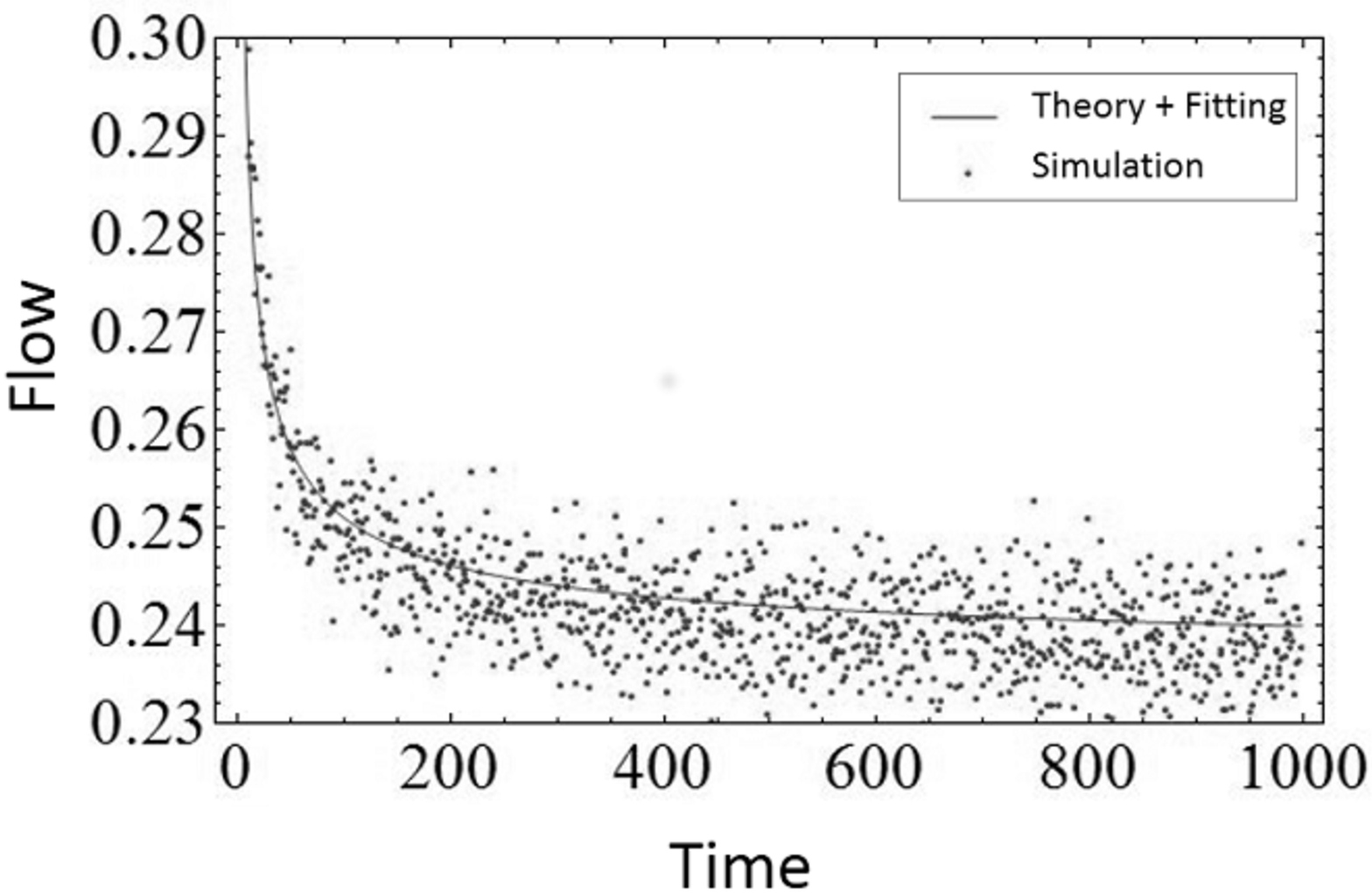}
\end{center}
\caption{Simulation results of $J_{n}-J_{n-1}$ are dotted against time steps after the state in the crossing cell opens for the parameter $p=0.72$. The value represents the average probability that vehicles exit the system at the time step. The solid line represents the result of the fitting using data of $J_{n}$ for $n=1,\ldots, 6$ based on theoretical analysis. The fitting function is $J_{n}^{fit} =  (1-\sqrt{1-p})/2 \cdot n + a n^{b}$. In this case, $a=0.39$ and $b=0.46$.}
\label{fig:1}
\end{figure}

In this subsection, we discuss the other approximation method: {\em isolated rarefaction wave approximation} (IRA). A schematic representation of the approximation method is shown in Fig.~\ref{fig:15}. In the HD phase, most lattice sites near the crossing cell are occupied. When the crossing cell opens, the cluster of vehicles exits the system. Simultaneously, the rarefaction wave propagates backward. Then, a pedestrian enters and the crossing cell turns into the closed state. If $\mu$ is small, it takes a large amount of time for the crossing cell to open again once it is closed, which enables the lattice sites near the crossing cell to become packed with vehicles. When the crossing cell opens again, the road area that is not packed with vehicles because of the rarefaction wave is located far from the crossing cell and it does not affect the flow of vehicles near the crossing cell. Moreover, the two rarefaction waves do not crash into each other. Thus we can neglect the effect of old rarefaction waves when considering the flow at the crossing cell. Approximating the system in this way, we can assume that all lattice cells are occupied whenever the crossing cell turns into the open state and treat the flow as a summation of the outflow of the cluster packed near the crossing cell. We call it IRA. We consider the number of time steps for which the crossing cell remains open once all the pedestrians exit. The conditional probability that the crossing cell remains open for $n$ steps when the crossing cell is open is $\left(\bar{\beta}/p\right)^{(n-1)\mu} \left(1-\left(\bar{\beta}/p\right)^\mu \right)$, where $\bar{\beta} =p e^{-\lambda/\mu}$. Since the mean probability of the open state is $\bar{\beta}/p$, the probability that the crossing cell remains open for $n$ time steps is given by $\left(\bar{\beta}/p\right)^{1+n\mu-\mu} \left(1-\left(\bar{\beta}/p\right)^\mu \right)^2$. Let $J_{n}$ denotes the summation of the outflow of the cluster when the crossing cell opens for $n$ steps in the case that all lattice sites are occupied at first. We obtain the flow 
\begin{equation} 
J_{HD}= \sum_{n=1}^{\infty}\left(\frac{\bar{\beta}}{p}\right)^{1+n\mu-\mu} \left(1-\left(\frac{\bar{\beta}}{p}\right)^\mu \right)^2 J_{n}.
\end{equation}
The summation of the outflow, $J_{n}$, is a function of $p$, but $J_{n}$ is unknown for all $n$, which indicates that $J_{n}$ requires an approximation. Although the flow at each time step $J_{n}-J_{n-1}$ is comparatively large when $n$ is even and otherwise it is small, $J_{n}-J_{n-1}$ is a decreasing function in a macroscopic view and converges to $(1-\sqrt{1-p})/2$ in the limit $n \to \infty$. Thus, we obtain $J_{n}$ as $(1-\sqrt{1-p})/2 \cdot n + o(n)$ and $J_{n}$ satisfies Eq.~(\ref{eq:mu0}), when $\bar{\beta}$ is finite, which is consistent with the exact limiting case. We calculate $J_{n}$ for $n=1,\ldots, 6$, considering the TASEP with six lattice sites and fit a curve to them. The function 
\begin{equation}
J_{n}^{fit} =  \frac{1-\sqrt{1-p}}{2} n + a n^{b}
\end{equation}
is used for curve fitting, where $a$ and $b$~($<1$) are parameters. In the case of $p=0.72$, we obtain the fitting parameters $a=0.39$ and $b=0.46$. The comparison between fitting results and simulation results is shown in Fig.~\ref{fig:1}. For a better illustration, $J_{n}-J_{n-1}$ is compared. This curve fitting agrees well with $J_{n}$ even for large $n$. Thus, this fitting is a good approximation of $J_{n}$. Then, the approximate flow is given using $J_{n}^{fit}$. The approximation results are shown in Figs.~\ref{fig:4}, \ref{fig:2}, \ref{fig:10}, and \ref{fig:7}(b). When $\bar{\beta}$ is large, the approximation results are larger than the simulation results because they neglect the effect of rarefaction waves bumping into each other. For small $\mu$, the approximation agrees well with the simulation, which indicates that the approximation scheme is valid.

\section{WITH a TRAFFIC LIGHT} 
\label{sec:6}

\subsection{Model description} 

\begin{figure*}
\begin{center}
\includegraphics[width=17cm]{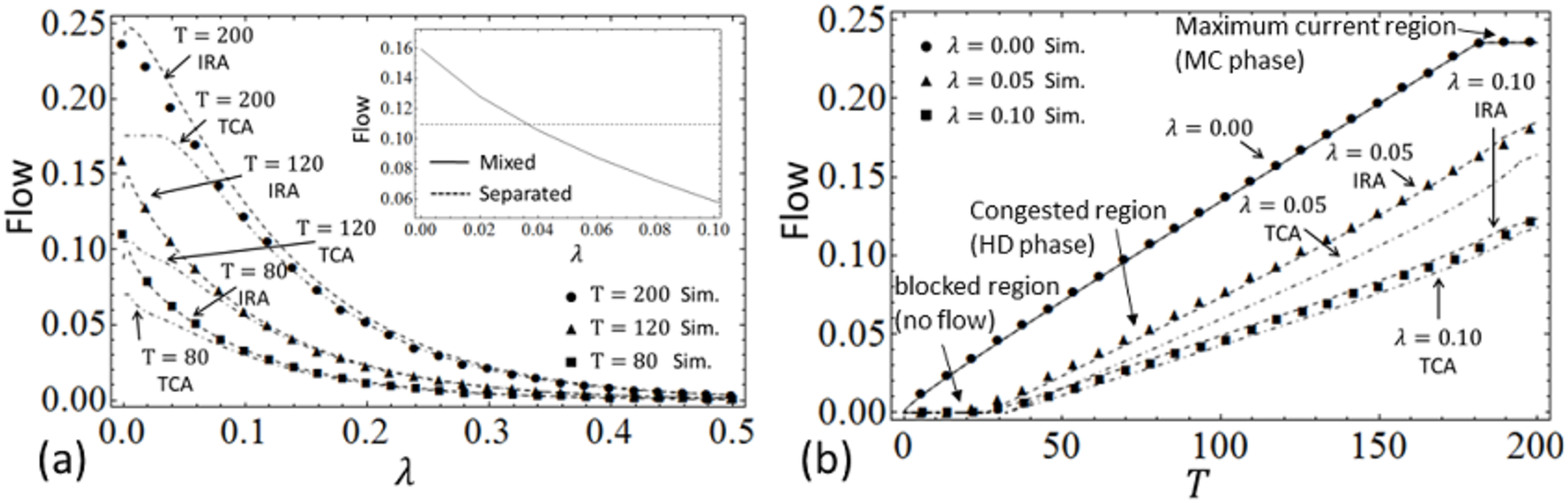}
\end{center}
\caption{(a) The average flow of vehicles against $\lambda$ for parameters $p=0.72$, $\mu=0.1$, $\alpha=1$, and $T+T'=200$. The circles, triangles and squares correspond to $T=200$, $120$, and $80$, respectively. The dot-dashed and dashed lines represent border-lines obtained by the TCA and IRA, respectively. The case of $T=200$ is the same as the case of no traffic lights. The supplemental figure shows the comparison between the pedestrian--vehicle separation signal and the ordinary signal (mixed). The green duration, $T$, of the separation signal is $80$ and that of the mixed signal is $120$. Forty time steps are devoted to pedestrians only. (b) The average flow against $T$ for parameters $p=0.72$, $\mu=0.1$, $\alpha=1$. The circles, triangles and squares correspond to $\lambda=0$ (no pedestrians), $0.05$, and $0.1$, respectively. The dot-dashed and dashed lines represent border-lines obtained by the TCA and IRA.}
\label{fig:7}
\end{figure*}

\begin{figure}
\begin{center}
\includegraphics[width=8cm]{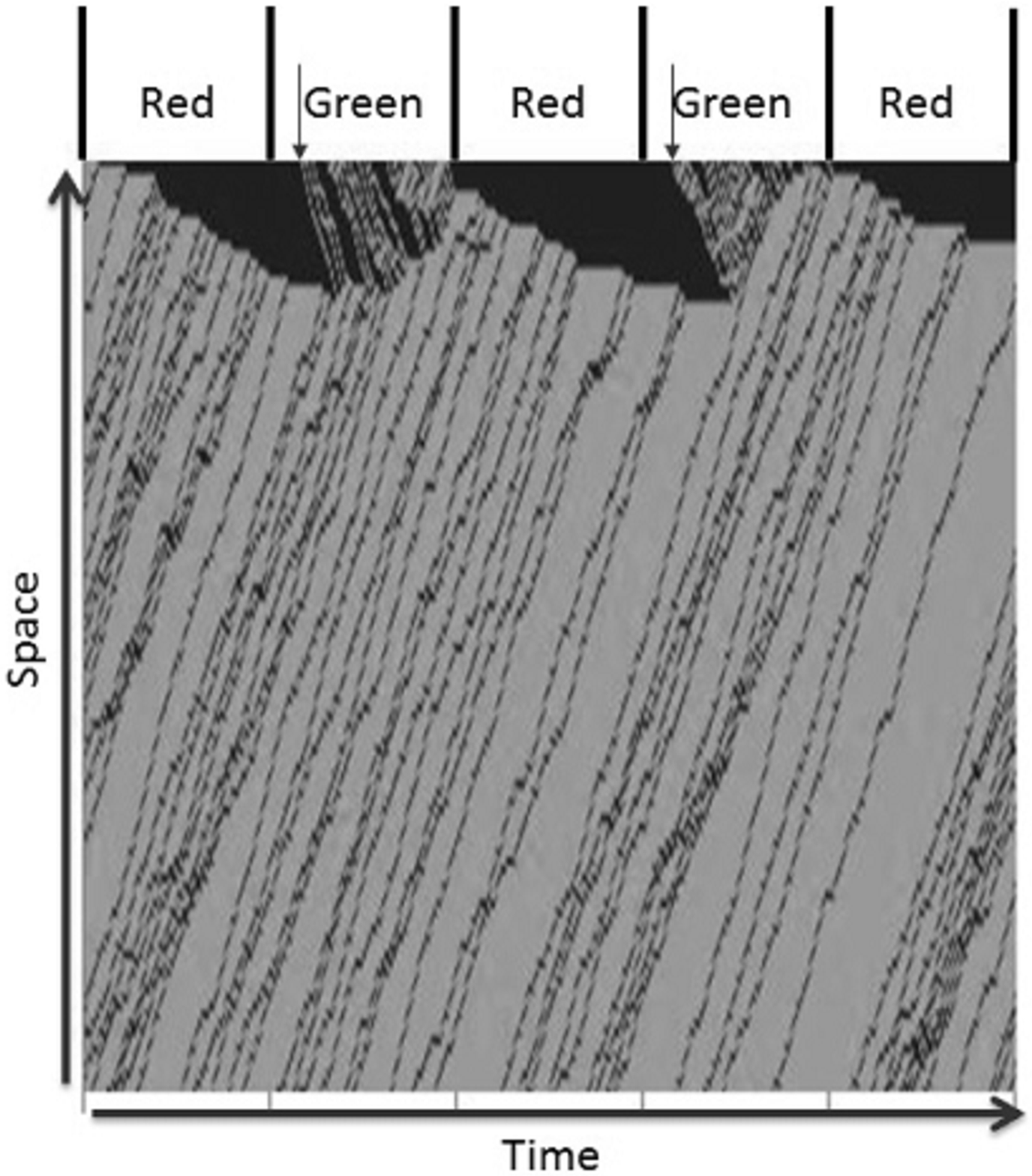}
\end{center}
\caption{Space-time plot of vehicles for $T=100$, $T'=100$, $\mu=0.2$, $\lambda=0.2$, and $\alpha=0.1$. In this figure, we set $N=110$ for a better illustration. The last $5 \times 10^2$ time steps out of $5 \times 10^5$ time steps in the simulation are used. The black and gray dots represent occupied and empty lattice sites, respectively. Each vehicle moves toward the upper right direction in the figure. The crossing cell is located at the upper side. Vehicles cannot exit when the light is red or pedestrians occupy the crossing cell. The color of the traffic light is shown above the figure. The arrows represent the time the first vehicle passes the crossing cell after the light turns green.}
\label{fig:6}
\end{figure}

In this section, we discuss the situation where vehicles try to turn right at an intersection with a traffic light. We assess the traffic capacity of intersections and compare it with the pedestrian--vehicle separation signals such as pedestrian scrambles. Figure~\ref{fig:13}(c) shows the schematic representation of the model when the traffic light is red. In real traffic, pedestrians and vehicles follow pedestrian signals and vehicles signals, respectively. Pedestrian signals turn red before vehicle signals turn red in Japan, which takes a middle position between two limiting cases: fully mixed (ordinary) traffic flow and fully segregated traffic flow such as pedestrian crossings. However, in this model, both vehicles and pedestrians follow the same traffic light to eliminate the time that only cars can turn because it is better to compare two limiting cases. When the traffic light is red, both vehicles and pedestrians cannot exit the system. Furthermore, vehicles and pedestrians whose direction is perpendicular to the road for our model can proceed then. The inflow of vehicles and pedestrians is the same as the previous model. We set the sum of the green and red durations to be $200$ time steps. The parameters $T$ and $T'$ denote the green and red durations, respectively. Other conditions are the same as the previous sections.

\subsection{Simulation results} 

Figure~\ref{fig:7} represents the traffic capacity of the flow with a traffic light and pedestrians. The horizontal axes of Fig.~\ref{fig:7}(a) and \ref{fig:7}(b) are $\lambda$ and $T$, respectively. The space-time plot of the model is shown in Fig.~\ref{fig:6}. The traffic capacity is reduced because of the traffic light, while the shape of the graph is not very different. Three regions, namely, the blocked region (no flow), congested region (HD phase) and maximum current region (MC phase), appear in this case. When the inflow of vehicles is not large, the free flow region (LD phase) appears as the case of the ordinary boundary condition. In the blocked region, pedestrians block traffic flow, and the flow is zero. It takes several time steps for vehicles to start moving again after the traffic light turns green. As $\lambda$ increases, the area of the blocked region becomes large. The traffic capacities of the pedestrian--vehicle separation signal and the ordinary (mixed) signal are shown in the supplemental figure of Fig.~\ref{fig:7}(a). The pedestrian--vehicle separation signal has a large traffic capacity for a wide range of $\lambda$ compared with the ordinary intersection. While the traffic capacity falls when there are pedestrians, the traffic light does not significantly cut down traffic flow.

\subsection{Theoretical analysis} 

\begin{figure}
\begin{center}
\includegraphics[width=8cm]{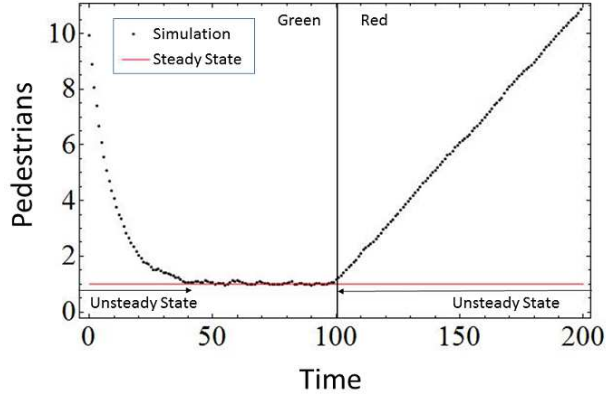}
\end{center}
\caption{(Color online) The average number of pedestrians in the crossing cell is plotted against time for parameters $\mu=0.1$, and $\lambda=0.1$. At time $t=0$ and $t=100$, the light turns green and red, respectively. The red line represents the theoretical results of the steady state.}
\label{fig:9}
\end{figure}

\begin{figure}
\begin{center}
\includegraphics[width=8cm]{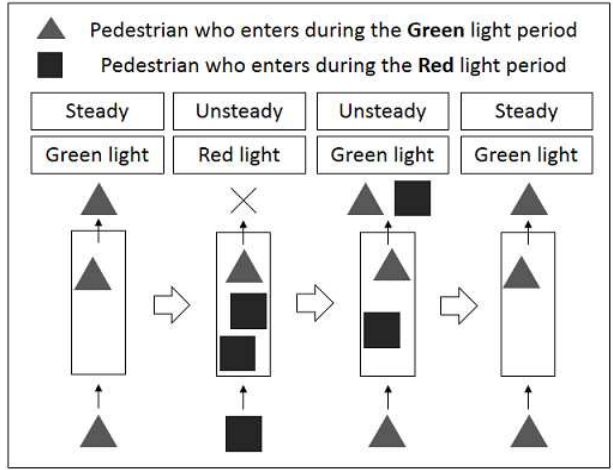}
\end{center}
\caption{Schematic representation of the pedestrian flow with a traffic light. We segregate two types of pedestrians. The triangular particles and square particles represent pedestrians who enter the crossing cell when the light is green and red, respectively. The triangular particles are in equilibrium at any time. When there is at least one square particle in the crossing cell, the flow is unsteady. The first figure and the last figure are the same.}
\label{fig:16}
\end{figure}

The traffic light makes the traffic and the pedestrian flows unsteady. First, we consider the pedestrian flow. The transition of the average number of pedestrians in the crossing cell against time is shown in Fig.~\ref{fig:9}. Time $0$ corresponds to the time when the traffic light turns green. Both the green light period and the red light period are $100$ time steps. The schematic representation of the transition of pedestrian flow is shown in Fig.~\ref{fig:16}. During the red light period, the number of pedestrians is increasing and the state is unstable. When the traffic light turns green, these pedestrians begin to simultaneously pass the crossing cell, and the crossing cell becomes crowded, which completely blocks traffic flow. After several time steps, a large number of pedestrians finish passing the crossing cell and the state becomes steady again. The pedestrian flow repeats the process. Then, we assess the unsteady time. Since all pedestrians are independent of other pedestrians, we segregate those who enter during the red light period. Other pedestrians, who enter during the green light period, are in equilibrium between the inflow and outflow at every time step because they do not enter or exit the system. Thus, the flow of other pedestrians remains steady even during the red light period. Assuming that segregated pedestrians successfully pass the crossing cell during the next green light period, the unsteady time is given by the time the segregated pedestrians stay in the system. We calculate the mean of the unsteady time $T_{unsteady}$. During the red light period, the average number of pedestrians who enter the crossing cell is $T'\lambda$. The approximate result of the average unsteady time is given by the maximum of  $T'\lambda$ i.i.d. geometric random variables $E(M_{T'\lambda}^{*})$, if $n$ is an integer. It is known that 
\begin{equation}
-\frac{1}{\ln{(1-\mu)}}  \sum_{k=1}^{n} \frac{1}{k} \le E(M_{n}^{*}) < -\frac{1}{\ln{(1-\mu)}} \sum_{k=1}^{n} \frac{1}{k}+1.
\end{equation}
for all $n$ \cite{eisenberg2008expectation}. Then, it is also easily seen that  
\begin{equation}
\sum_{k=1}^{n} \frac{1}{k} \approx \ln{n}+\gamma + \frac{1}{2n}-\sum_{k=1}^{\infty}\frac{B_{2k}}{2kn^{2k}},
\end{equation}
where $\gamma = 0.57\ldots$ is the Euler-Mascheroni constant and $B_{k}$ are Bernoulli numbers. Thus, we obtain 
\begin{equation}
T_{unsteady} \approx -\frac{1}{\ln{(1-\mu)}} \left(\ln{T' \lambda}+\gamma+ \frac{1}{2n}-\sum_{k=1}^{\infty} \frac{B_{2k}}{2kn^{2k}}\right).
\end{equation}
The domain of the definition of the number of pedestrians is extended to real numbers as an approximation. Additionally, the term $1/2n-\sum_{k=1}^{\infty}B_{2k}/2kn^{2k}$ is neglected for the sake of simplicity. If $T_{unsteady}>T$, the traffic light turns red again before all packed pedestrians pass the crossing cell, which leads to the blocked state. In the case of $T_{unsteady}<0$, we assume there are no unsteady time steps.

Next, we discuss traffic flow. We only discuss the HD phase because the flow is independent of the crossing cell in other phases. In the unsteady time of pedestrian flow, no vehicles can exit the system. It takes several time steps for the first vehicle to pass the crossing cell after the light turns green because of the unsteady state of pedestrians flow. Subtracting the unsteady time from the green duration, $T$, the substantial green duration for traffic flow is given by
\begin{equation} 
T_{sub} \approx T+\frac{1}{{\ln{(1-\mu)}}}(\ln{T' \lambda}+\gamma).
\end{equation} 
If $T_{sub}<0$, we assume $T_{sub}=0$ and if $T_{sub}>T$, we assume $T_{sub}=T$. In the HD phase, the road near the crossing cell is packed with vehicles when the traffic light turns green. The outflow is $J_{T_{sub}}$ per one cycle of the traffic light in the case where no pedestrians obstruct traffic flow, i.e., $\lambda=0$. We obtain simpler approximate results for the outflow if many pedestrians interfere with traffic flow. The effect of pedestrians on traffic flow is dominant compared with that of the traffic light. With sufficient numbers of pedestrians, they cut the cluster of traffic flow into many clusters. The flow is weakly dependent on the time steps after the traffic light turns green. Thus, the flow is approximately proportional to the substantial green duration $T_{sub}$ in this case. Therefore, we obtain the flow
\begin{eqnarray}
  J_{HD} = \begin{cases}
    \displaystyle \frac{T_{sub}}{T} p (P_{0,1}^{0}+P_{1,1}^{0}) & (TCA) \\
    \displaystyle \frac{T_{sub}}{T} \sum_{n=1}^{\infty} \left(\frac{\bar{\beta}}{p}\right)^{1+n\mu-\mu} \left(1-\left(\frac{\bar{\beta}}{p}\right)^\mu \right)^2 J_{n} & (IRA). \nonumber
  \end{cases}
\end{eqnarray}
Figure~\ref{fig:7} compares the simulation and theoretical results. The result of $\lambda=0$ is plotted using the method shown above. For $\lambda \neq 0$, the results of the theoretical analysis are shown in dot-dashed lines (with the TCA) and dashed lines (with the IRA). The theoretical results are well in agreement with the simulation results.

\section{CONCLUSION} 
\label{sec:7}

In this paper, we studied the bottleneck situation of traffic flow by considering the TASEP with interaction among vehicles turning right, pedestrians, and a traffic light. We found that pedestrians passing the crossing cell reduce the traffic capacity more than the ordinary TASEP with the same $\bar{\beta}$. The effect of the interaction is much larger in cases where pedestrians' walking speed $\mu$ is small. Moreover, we investigated two solvable limiting cases: $\mu=1$ and $\mu \rightarrow 0$. Additionally, we proposed two types of approximation: (2+1)-cluster approximation (TCA) and isolated rarefaction wave approximation (IRA). The result of the former is in agreement with the simulation results when $\mu$ is sufficiently large. The result of the latter captures the essential features of the flow when $\mu$ is sufficiently small. Moreover, with a traffic light, we studied the phenomenon that pedestrians waiting during the red light period prevent vehicles from passing the crossing cell until a few time steps after the light turns green. We theoretically obtained the approximate average duration of the phenomenon. We compared the pedestrian--vehicle separation signal such as the pedestrian crossing and the ordinary pedestrian--vehicle mixed signal. The result shows that the traffic capacity of the ordinary signal with even few pedestrians is smaller than that of the pedestrian--vehicle separation signal. The pedestrian--vehicle separation signal is a better choice if $\lambda > 0.037$, according to the supplemental figure of Fig.~\ref{fig:7}(a). The effect of pedestrians blocking the crossing is stronger than that of the traffic light in this case. If the lengths of one cell, one time step, $\mu$, and the hopping probability, $p$, are set as $7.5$ m, $0.54$ s, $0.1$, and $0.72$, respectively, the average time pedestrians take to pass the crossing cell is $5.4$ s and the border value corresponds to approximately $4.1$ pedestrians per minute. Since the effect of pedestrians is huge, this amount is very small. Furthermore, we can identify an exponential decrease in flow by pedestrians. The sharp drop in flow packs the right turn lane with vehicles, which leads to terrible congestion. The result surely indicates that a pedestrian crossing is a good way to prevent traffic jam.

\section*{ACKNOWLEDGMENTS}
The author would like to thank Takumi Masuda for helpful discussions and suggestions.

\section*{APPENDIX: DERIVATION OF STEADY PEDESTRIAN FLOW}
\renewcommand{\theequation}{A.\arabic{equation}}
\setcounter{equation}{0}

In this section, we derive the features of pedestrian flow in the steady state. We define the distribution of the number of pedestrians $v={}^t\!(\pi_0,\pi_1,\pi_2,\ldots)$, where $\pi_{n}$ denotes the probability that there are $n$ pedestrians in the crossing cell. The transition matrix is given by $T_{cro} = (P^{i,j})$ $(i,j = 0,1,2,\ldots)$ where
\begin{equation}
P^{i,j} = \sum_{k=0}^{min(i,j)} {j \choose k} (1-\mu)^{k} \mu^{j-k} \frac{\lambda^{i-k}}{(i-k)!} e^{-\lambda}.
\end{equation}
We emphasize that more than one pedestrian may enter or exit the crossing cell during one discrete time step, unlike a M/M/$\infty$ queue, which is defined in continuous time. The stationary state of the master equation is given by $v=T_{cro} v$. We solve the equation using a generating function. Comparing the Taylor series expansion of functions, we obtain
\begin{eqnarray}
\sum_{n=0}^{\infty} P^{i,j} z^n &=& \sum_{k=1}^{j} {j \choose k} (1-\mu)^{j-k} \mu^k z^{j-k} e^{-\lambda(z-1)} \nonumber \\
&=& \left[\mu+(1-\mu)z\right]^j e^{-\lambda(z-1)}.
\end{eqnarray}
Thus, using $v=T_{cro} v$, the probability generating function of $v$ is given by
\begin{equation}
p(z) = \sum_{k=0}^{\infty} \pi_{k} z^k = \sum_{j=0}^{\infty} \pi_{j} \left[\mu+(1-\mu)z\right]^j e^{-\lambda(z-1)}.
\label{generating}
\end{equation}
Considering the steady distribution of the number of pedestrians in the case of the M/M/$\infty$ queue is the Poisson distribution, we assume that the distribution in this case is also the Poisson distribution, $\pi_{n} = 1/n! \cdot a^n e^{-a}$, with parameter $a$. Substituting $\pi_{n}$ for (\ref{generating}), we confirm that the equation is satisfied if $a=\lambda/\mu$. As a result, we obtain
\begin{equation}
\pi_{n} = \frac{1}{n!} \left( \frac{\lambda}{\mu}\right)^n e^{-\frac{\lambda}{\mu}}.
\end{equation}
Additionally, the average of the number of pedestrians in the crossing cell in the steady state is given by $\lambda/\mu$.



\end{document}